\documentclass[aps,prl,preprint,superscriptaddress]{revtex4}
\usepackage{graphicx}

\def\Blue{} 
\def\Black{} 
\def\Red{} 
\begin{document}

\title{ Entwined Pairs and Schr{\"o}dinger 's Equation}
\author{G. N. Ord}
\email[ corresponding author ]{gord@acs.ryerson.ca}
\affiliation{M.P.C.S. \\
Ryerson University\\
Toronto Ont.}
\author{R.B. Mann}
\email{mann@avatar.uwaterloo.c}
\affiliation{Dept of Physics\\
University of Waterloo\\
Waterloo, Ont Canada}
\date{May 18 2002}

\begin{abstract}
We show that a point particle moving in space-time on entwined-pair paths
generates Schr\"{o}dinger's equation in a static potential in the
appropriate continuum linit. This provides a new realist context for the Schr%
\"{o}dinger equation within the domain of classical stochastic processes. It
also suggests that self-quantizing systems may provide considerable insight
into conventional quantum mechanics.
\end{abstract}

\date{June 14 2002}
\maketitle





\section{Introduction}

In quantum mechanics, the wavefunction is part of an algorithm that allows
us to predict the behaviour of a system. Although the predictive power of
the algorithm and the rules for the evolution of the wavefunction are not in
doubt, assigning an objective reality to the wavefunction is usually not
done. The wavefunction itself is not measurable, and there is no consensus
within the physics community as to the extent of its role beyond that of a
calculational device.

Because of the difficulty in interpreting the wavefunction in quantum
mechanics, there have been a series of attempts at constructing simple
classical statistical mechanical systems which have, as part of their
description, either the Schr{\"{o}}dinger or Dirac equations
\cite{gord92,gord93lett,gord96,gordDeakin97,McKeonOrd92,gord01b}. The idea
has been to extract the `quantum' equations in classical statistical models
without forcing an analytic continuation. This puts the relevant equation
in a context where the wave function solutions represent ensemble averages
of known microscopic quantities. The motivation for this program of study
stems from the fact that in conventional quantum mechanics, it is the
Formal Analytic Continuation (FAC) that brings about wave-particle duality.
However, the FAC implemented by canonical quantization tells us nothing
about how nature accomplishes the same task. Finding Classical systems that
imitate quantum propagation without FACs potentially suggests ways in which
nature could effect the analytic continuation purely through the geometry
of space-time trajectories. Consequently the pursuit of such model systems
offers the hope of obtaining a deeper understanding of quantum mechanics.

The latest such model is completely `self-quantizing' and uses only
space-time geometry to provide quantum interference effects usually produced
by FAC\cite{gord01a}. The self-quantization is accomplished by
`entwined-paths'. An entwined space-time path is one in which for each path
that extends from the origin $(z,t)=(0,0)$ to some return point $(z,t_{R})$, 
$t_{R}>0$, there is a return path from $(z,t_{R})$ to the origin that
oscillates about the original path in such a way that the combined pair
imitates Feynman Chessboard paths with the accompanying phase rule Fig.(1).
The Entwined model directly yields a 4-component Dirac equation for a
one-dimensional system \cite{gnorbm1}.

\begin{figure}[tbp]
\includegraphics[scale = .42]{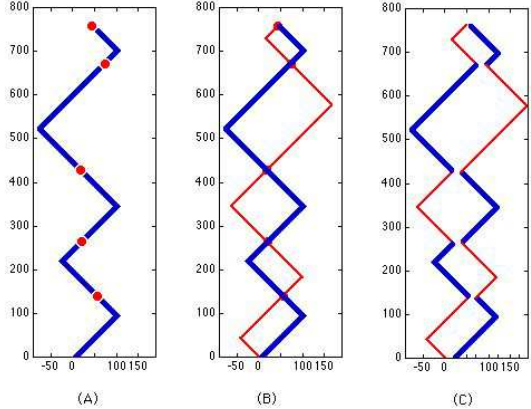}
\caption{Forming entwined paths in space-time: $z$ is horizontal, $t$ is
vertical. The particle travels at constant speed but occasionally reverses
direction in response to a stochastic process. (A) A stutter is introduced
into the stochastic process. At every other indication from the
stochastic process, a marker is dropped instead of a direction change (red
disk in fig). After some specified time $t_{R}$, stop at the next marker.
(B) Reverse direction in time but not in space. Follow the `light-cone'
paths through the markers back to the origin. (C) The entwined path formed
in (B) can be regarded as two osculating paths which we call envelopes.
These are separated in (C) for clarity. The geometry of the envelopes is the
same as if the paths were generated by the stochastic process without a
stutter. The colouring denoting direction of traversal mimicks the
Chessboard colouring.}
\label{entwined}
\end{figure}
In this paper we demonstrate the connection between the entwined model and
Schr\"{o}dinger's equation in two ways. First we use a technique developed
in \cite{gordDeakin96} to obtain Schr\"{o}dinger's equation for a particle
in a static potential, as the direct continuum limit of the original
discrete model. This method mimics the usual transition from a symmetric
random walk to the diffusion equation. However the use of entwined paths
ensures that the resulting phenomenology is reversible, and the result is
Schr\"{o}dinger 's equation.

In the second method we take the continuum limit maintaining both a finite
characteristic length (mean free path) and a finite characteristic speed.
This version mimics Kac's derivation of the Telegraph
equations\cite{Kac74,Kac79}. Here the entwined paths again ensure
reversibility and the result is a particular representation of the Dirac
equation. The subsequent limit in which the mean free speed goes to
infinity yields Schr\"{o}dinger 's equation.

\section{Entwined pairs: a Classical Stochastic Model}

Consider the following stochastic process (Fig.(\ref{entwined})) . A single
particle is constrained to move in discrete time on a lattice with lattice
spacing $\delta $. The time steps are of length $\epsilon $. At each step
the particle moves one lattice spacing in $z$ and one in $t$. For a point
source, the particle starts at the origin at $(0,0)$ and steps to the first
lattice point at $(\delta ,\epsilon )$. The particle then steps to $(2\delta
,2\epsilon )$ with probability $\beta $ or changes direction and steps to $%
(0,2\epsilon )$ with probability $\alpha $. Once it is moving in the $-z$
direction the particle first drops a marker with probability $\alpha $ and
then at the second indication of the stochastic process changes direction
again. This alternating sequence of changing direction and dropping markers
is continued until the first marker after some specified return time $t_{R}$%
. At that marker the particle maintains its direction in space but reverses
direction in time. Subsequent steps of the process follow the light cones of
the markers back to the origin. At the origin the process begins again.

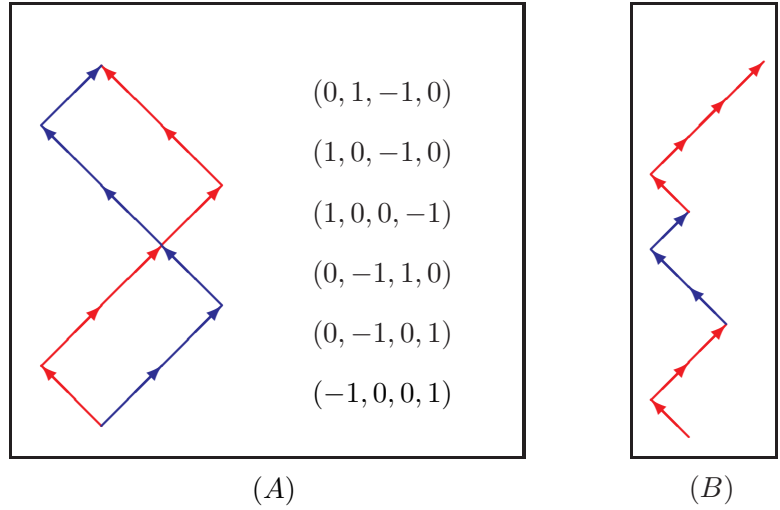
\begin{figure}[tbp]
\setlength{\unitlength}{.8mm} 
\begin{picture}(70,80)(-25,-5)
\put(45 ,4){$( -1,0, 0,1)$}
\put(35 ,-12){$(A)$}
\thicklines
\put(10,0){\Red{\vector(-1,1){10}}}
\put(10,0){\Blue{\vector(1,1){10}}}\Black
\put(45,14){$( 0,-1,0,1)$} 
\put(0,10){\Red{\vector(1,1){10}}}
\put(20,10){\Blue{\vector(1,1){10}}}\Black
\put(45,24){$(0,-1,\Black 1,0)$} 
\put(10,20){\Red{\vector(1,1){10}}}
\put(30,20){\Blue{\vector(-1,1){10}}}\Black
\put(45,34){$( 1,0,\Black 0,-1)$} 
\put(20,30){\Red{\vector(1,1){10}}}
\put(20,30){\Blue{\vector(-1,1){10}}}\Black
\put(45,44){$( 1,0,\Black -1,0)$} 
\put(30,40){\Red{\vector(-1,1){10}}}
\put(10,40){\Blue{\vector(-1,1){10}}}\Black
\put(45,54){$( 0,1,\Black -1,0)$} 
\put(20,50){\Red{\vector(-1,1){10}}}
\put(0,50){\Blue{\vector(1,1){10}}}\Black
\put(-5,-5){\framebox(85,75)}
\end{picture}\hspace{3cm} \setlength{\unitlength}{.5mm} 
\begin{picture}(70,80)(-25,-5)
\put(10 ,-16){$(B)$}
\thicklines
\put(10,0){\Red{\vector(-1,1){10}}}
\put(0,10){\Red{\vector(1,1){10}}}
\put(10,20){\Red{\vector(1,1){10}}}
\put(20,30){\Blue{\vector(-1,1){10}}}\Black
\put(10,40){\Blue{\vector(-1,1){10}}}\Black
\put(0,50){\Blue{\vector(1,1){10}}}
\put(10,60){\Red{\vector(-1,1){10}}}
\put(0,70){\Red{\vector(1,1){10}}}\Black
\put(10,80){\Red{\vector(1,1){10}}}
\put(20,90){\Red{\vector(1,1){10}}}\Black
\put(-5,-5){\framebox(38,120)}
\end{picture}
\caption{A) A sample entwined path with the corresponding velocity 4-vector.
Note the change in sign of the two envelopes when the particles cross. (B) A
left envelope path. Note that the left envelope path changes colour at every
left-hand corner.}
\label{states}
\end{figure}

In the figure, suppose the portion of the trajectory traversed forward in
time is coloured blue, with the backward portion coloured red. Whereas the
colours we have chosen are arbitrary, the fact that we distinguish the
direction of traversal in time is not. Since a particle that reverses its
direction in time will be seen to be anihilated by an `antiparticle' which
is really the same particle reversing its direction in time, we shall define
a `charge' associated with the traversal of a trajectory. Forward traversals
will correspond to a charge of $+1$, reversed traversals will correspond to
a charge of $-1$. In the figure, blue portions of the trajectory correspond
to a charge of $+1$ and red portions correspond to $-1$. Note that although
each complete entwined path yeilds a net charge of zero at any value of $t$,
entwined paths do separate charge and we shall see that the charge
separation builds up a charge field in space-time that is oscillatory in
nature. We can then imagine counting the net charge which enters a site.
Regarding the entwined pair as a coherent physical entity we will have a
four component object to consider, since there are two paths to each
entwined pair, and each member of the pair has two possible directions to
move in space Fig.(\ref{states} A). 

We have generated the entwined pair in a way which allows us to see how it
is equivalent to a single closed loop in space-time. However the method of
generation is tuned to counting the paths based on their outer envelopes.
That is, both the left and right `corners' in the outer envelopes are
generated statistically by the same stochastic process. When we dropped
markers at every other call from the random process, and then passed through
the markers on the return path, we ensured that we could equally well have
generated the outer envelopes by an alternating colouring of an
independently generated outer envelope. For example, the left envelope in
Figure(\ref{states}B) may be obtained by starting on the red path at the
origin and alternating blue and red sections at every second envelope
corner. All left envelope paths have this colouring rule, which itself is a
consequence by the geometry of entwined pairs. Note that on both envelopes,
the probability of a corner is always $\alpha $ regardless of whether it is
a right or left-handed corner.

\begin{figure}[tbh]
\setlength{\unitlength}{.8mm} 
\begin{picture}(70,80)(-25,-5)

\thicklines
\put(10,0){\Red{\vector(-1,1){10}}}
\put(-11,-1){\Blue{\vector(1,1){10}}}\Black
\put(00,10){\Red{\vector(-1,1){10}}}
\put(-1,9){\Red{\vector(-1,1){10}}}\Black
\put(25,10){Contributions to $\phi_1^{\prime }$} 
\put(10,28){\Red{\vector(-1,1){10}}}
\put(-11,29){\Red{\vector(1,1){10}}}\Black
\put(00,38){\Red{\vector(1,1){10}}}
\put(-1,39){\Red{\vector(1,1){10}}}\Black
\put(25,40){Contributions to $\phi_2^{\prime }$} 

%
\put(-15,-5){\framebox(85,65)}
\end{picture}\hspace{3cm}
\caption{$\protect\phi_1^{\prime }$ and $\protect\phi_2^{\prime }$ on the
left envelope receive contributions from the previous time step. Note that $%
\protect\phi_1^{\prime }$ changes colour at a left-hand corner. $\protect\phi%
_2^{\prime }$ does not change colour at a right-hand corner.}
\label{contrib}
\end{figure}
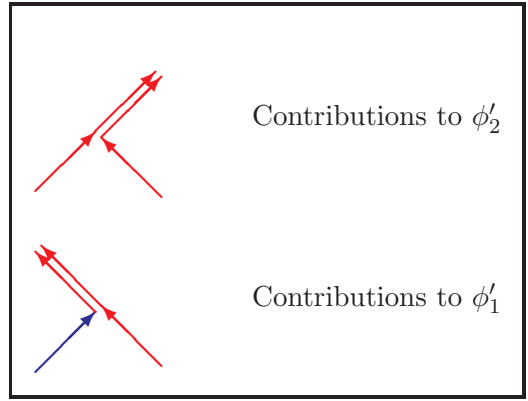

Regarding the left envelope, if $\phi_{1}^{\prime }(z,t+\epsilon )$ is the
net charge entering the lattice points $(z,t+\epsilon )$ from the $+z$
direction and $\phi _{2}^{\prime }(z,t+\epsilon )$ is the net charge
entering from the $-z$ direction then the difference equation expressing
conservation of charge is easily found. Regarding Fig.(\ref{contrib}) we see
that for the charge on the left envelope we have. 
\begin{eqnarray}
\phi _{1}^{\prime }(z,t+\epsilon ) &=&\beta \phi _{1}^{\prime }(z+\delta,t
)-\alpha \phi _{2}^{\prime }(z-\delta,t )  \label{diff1} \\
\phi _{2}^{\prime }(z,t+\epsilon ) &=&\beta \phi _{2}^{\prime }(z-\delta,t
)+\alpha \phi _{1}^{\prime }(z+\delta,t )  \nonumber
\end{eqnarray}
Note that as $\phi_2^{\prime}$ scatters into $\phi_1^{\prime}$ it changes
sign so the contribution to $\phi_1^{\prime}$ is negative.

Similarly, for the right envelope, if $\phi _{3}^{\prime }(z,t+\epsilon )$
is the net number entering the lattice points $(z,t+\epsilon )$ from the $+z$
direction and $\phi _{4}^{\prime }(z,t+\epsilon )$ is the net number
entering from the $-z$ direction we may write:

\begin{eqnarray}
\phi _{3}^{\prime }(z,t+\epsilon ) &=&\beta \phi _{3}^{\prime }(z+\delta,t
)+\alpha \phi _{4}^{\prime }(z-\delta,t )  \nonumber \\
\phi _{4}^{\prime }(z,t+\epsilon ) &=&\beta \phi _{4}^{\prime }(z-\delta,t
)-\alpha \phi _{3}^{\prime }(z+\delta,t )  \label{diff2} \\
&&  \nonumber
\end{eqnarray}

Note the alternating signs in the `scattering' terms in these equations.
These arise because of the fact that every other `corner' in an envelope is
actually an exchange with an antiparticle. The two envelopes naturally
partition the states into two block diagonal systems coupled only by their
initial conditions. Versions of these equations have been obtained and
numerically verified for the underlying stochastic process in \cite{gnorbm1}.

\section{Entwined Pairs and Diffusive Scaling}

Symmetric binary random walks have a well-known scaling that allows a
description in the continuum limit. This scaling is termed diffusive scaling
and corresponds to 
\begin{equation}
(\delta )^{2}/(2\epsilon )\,\rightarrow D\qquad \mathrm{{as}\qquad }\epsilon
\rightarrow 0  \label{limit}
\end{equation}%
where $D>0$ is a diffusion constant. The reason that this scaling `works' is
simply because the \emph{mean-square} end-to-end length of a symmetric
binary random walk increases as the number of steps $n$, so (\ref{limit}) is
the natural scaling for such walks.

In the Entwined pair model above, if we choose $\alpha $ and $\beta $ to be
asymptotically $1/2$ then the underlying envelopes are asymptotically
symmetric binary random walks. However the entwining of paths makes each
envelope with its orthogonal twin a reversible path. The result is that a
phenomenology that would otherwise be described by the diffusion equation is
instead described by the Schr{\"{o}}dinger equation, reflecting the change
from an entropy dominated to an entropyless system. The new system is
entropyless because each time-reversed path essentially undoes the disorder
from its original partner .

To proceed, we scale system (\ref{diff1}) diffusively. That is, we shall let 
$\alpha $ and $\beta $ approach $1/2$ as the lattice spacing gets small. The
potential will then enter through the limiting process in just how the terms 
$\alpha $ and $\beta $ approach $1/2$.

Consider the left envelope densities $\phi^{\prime}_1$ and $\phi^{\prime}_2$%
. We will write (\ref{diff1}) in matrix form using shift operators $%
E^{\pm}_z $ and $E_t$ where $E^{\pm}_z \phi^{\prime}(z,t)=\phi^{\prime}(z\pm
\delta,t)$ and $E_t \phi^{\prime}(z,t)=\phi^{\prime}(z,t+\epsilon)$. Writing 
$\Phi^{\prime}=\left ( 
\begin{array}{c}
\phi^{\prime}_1 \\ 
\phi^{\prime}_2%
\end{array}
\right) $, (\ref{diff1}) may be written 
\begin{equation}
E_t \Phi^{\prime}(z,t)=\left ( 
\begin{array}{cc}
\beta E_z^+ & -\alpha E_z^- \\ 
\alpha E_z^+ & \beta  E_z^-%
\end{array}
\right) \Phi^{\prime}(z,t).  \label{mat1}
\end{equation}
Now suppose the particle chooses its next step according to a canonical
ensemble in which a smooth bounded potential $v(z)\epsilon$ acts like an
energy. That is, suppose the particle associates a relative energy of $%
+v(z)\epsilon$ to continue the direction of travel, and an energy of $%
-v(z)\epsilon$ for a direction reversal. For large positive values of $v(x)$
reversals are favoured. This means that the reversal probability is

\begin{equation}
\alpha =\frac{e^{v(z)\epsilon }}{e^{-v(z)\epsilon }+e^{v(z)\epsilon }}
\label{alf}
\end{equation}%
so that 
\begin{equation}
\alpha =\frac{1}{2}(1+v(z)\epsilon )+O(\epsilon ^{2}),\mathrm{\quad \ \quad }%
\beta =\frac{1}{2}(1-v(z)\epsilon )+O(\epsilon ^{2})
\end{equation}%
Note that the effect of the field is to alter the local mean free path of
the particle. If $v(z)$ is negative the particle tends to stay moving in the
same direction for longer periods. Conversely, if $v(z)$ is positive the
particle changes direction more frequently.

Now we wish to approximate solutions of (\ref{mat1}) for small $\delta $ by
solutions of a partial differential equation. To do this we expand the shift
operators 
\begin{equation}
E_{z}^{\pm 1}=1\pm \delta \frac{\partial }{\partial z}+\frac{1}{2}\delta ^{2}%
\frac{\partial ^{2}}{\partial {z}^{2}}+O(\delta ^{3}).  \label{shiftx}
\end{equation}%
and 
\begin{equation}
E_{t}=1+\epsilon \frac{\partial }{\partial t}+O(\epsilon ^{2}).
\label{shiftt}
\end{equation}%
The matrix in (\ref{mat1}) may then be written: 
\begin{eqnarray}
\left( 
\begin{array}{cc}
\beta E_z^+ & -\alpha E_z^- \\ 
\alpha E_z^+ & \beta  E_z^-%
\end{array}
\right)  &=&\frac{1}{2}(1-v(z)\epsilon )(I+\sigma _{z}\delta \frac{\partial 
}{\partial z}+I\frac{1}{2}\delta ^{2}\frac{\partial ^{2}}{\partial {z}^{2}})
\\
&+&\frac{1}{2}(1+v(z)\epsilon )(\sigma _{q}+\sigma _{x}\delta \frac{\partial 
}{\partial z}+\sigma _{q}\frac{1}{2}\delta ^{2}\frac{\partial ^{2}}{\partial 
{z}^{2}})+O(\delta ^{3})
\end{eqnarray}%
where $\sigma _{x}$ and $\sigma _{z}$ are the usual Pauli matrices, $\sigma
_{q}$ is $-i\,\sigma _{y}$ and $I$ is the $2\times 2$ identity matrix.
Keeping only the lowest order terms this is 
\begin{eqnarray}
\left( 
\begin{array}{cc}
\beta E_z^+ & -\alpha E_z^- \\ 
\alpha E_z^+ & \beta  E_z^-%
\end{array}
\right)  &=&\frac{1}{2}(I+\sigma _{q})+\frac{1}{2}(\sigma _{z}+\sigma
_{x})\delta \frac{\partial }{\partial z}+\ldots   \nonumber \\
&+&\frac{1}{2}(I+\sigma _{q})(\delta ^{2}\frac{\partial ^{2}}{\partial {z}%
^{2}})-\frac{1}{2}(I-\sigma _{q})v(z)\epsilon   \label{expand} \\
&&  \nonumber
\end{eqnarray}%
The term in (\ref{expand}) that is independent of both $\delta $ and $%
\epsilon $ is an unnormalized finite rotation. If we are going to match
solutions of the difference equation to that of a differential equation we
must correct the normalization and the finite rotation. To correct the
normalization we write $\Phi (z,t)=(\sqrt{2})^{t/\epsilon }\Phi ^{\prime
}(z,t)$, and to avoid the finite rotation we note that $(\frac{1}{\sqrt{2}}%
(I+\sigma _{q}))^{8}=I$. As long as we restrict our comparison of the
difference equation and the differential equation so that steps in the $t$
direction are integer multiples of $8\epsilon $, we will avoid the finite
rotations, which are in any case an artifact of the diffusive
(non-relativistic) limit. The difference equation we are considering is then:

\begin{equation}
E_t^8 \Phi(z,t)=2^4\left ( 
\begin{array}{cc}
\beta E_+ & -\alpha E_- \\ 
\alpha E_+ & \beta E_-%
\end{array}
\right)^8 \Phi(z,t).  \label{mat2}
\end{equation}%
This is the original difference equation, transformed to remove a decaying
exponential, and viewed eight steps at a time. Expanding equation(\ref{mat2}%
), keeping lowest order terms and using (\ref{limit}) we get, after some
algebra:

\begin{equation}
\frac{\partial }{\partial t}\left[ 
\begin{array}{c}
\phi _{1} \\ 
\phi _{2}%
\end{array}%
\right] =\left[ 
\begin{array}{cc}
0 & D\frac{\partial ^{2}}{\partial z^{2}}-v(z) \\ 
-D\frac{\partial ^{2}}{\partial z^{2}}+v(z) & 0%
\end{array}%
\right] \left[ 
\begin{array}{c}
\phi _{1} \\ 
\phi _{2}%
\end{array}%
\right] +O(\delta )  \label{blockde}
\end{equation}

where the $\phi $ are real. We may express this in complex form as 
\begin{equation}
i\frac{\partial }{\partial t}(\phi _{2}+i\phi _{1})=(-D\frac{\partial ^{2}}{%
\partial z^{2}}+v(z))(\phi _{2}+i\phi _{1})+O(\delta ).  \label{schrod}
\end{equation}%
Thus solutions of Schr{\"{o}}dinger 's equation (\ref{schrod}) approximate
solutions of the entwined pair difference equation (\ref{mat2}) for small $%
\delta $. Note that $\phi _{1}$ and $\phi _{2}$ here are real functions
which are themselves just limits of ensemble averages of the net charge
accumulated via entwined paths. They are not components of wave functions in
the sense of quantum mechanics, since there has been no FAC or quantization
involved. The $\phi $ are strictly real classical objects representing an
expected net flow of charge in the $-z$ and $+z$ directions respectively.

We also note that the potential energy $v(z)$ that effected the particle's
mean free path in (\ref{alf}) enters the Schr{\"{o}}dinger equation as a
potential term in the Hamiltonian. Although $v(z)$ does not favour either
direction explicitly in the actual walk, as the particle moves it tends to
reverse direction more frequently when moving in a region of high potential
energy, and less frequently when moving in a region of low potential. The
net effect at the level of the Schr{\"{o}}dinger equation is that regions of
maximum $v(z)$ tend to repel the particle and regions of minimum $v(z)$ tend
to attract it.

The remaining two densities satisfy the time-reversed version of (\ref%
{blockde}) and are remnants of the four component description of entwined
pairs. The two PDE's are connected by their initial conditions and their
solutions, in complex form, are conjugates of each other, so their product,
integrated over all space is a time-independent positive constant. This fact
opens the door to the Born postulate which would associate a probability
density with the product of the two wavefunctions. However, at this point we
have no justification for invoking such a postulate. In future work we shall
examine the microscopic dynamics to see if the postulate is implied by any
reasonable measurement scheme.

Although in the above calculation we were forced to look at the densities
only every eight steps because of the finite rotation at each step, there is
an alternative approach. We could define a new set of densities which
rotated by $\pi /4$ with every step as the ensemble of walks actually does.
This would allow us to avoid the `stroboscope' approach above, and we could
take the continuum limit as an approximation of a single step. In the next
section we avoid the problem entirely by taking the continuum limit with a
finite signal velocity. This in turn changes the finite rotation at each
step into an infinitesimal rotation which admits a continuum limit directly.

\section{Entwined pairs with fixed signal velocity}

In the previous section the asymptotic scaling was diffusive. In the absence
of a potential field, at each lattice scale $(\delta ,\epsilon )$, the
probability that the envelope walk turns left or right is exactly $1/2$.
This symmetry reflects the paradigm that the walk is actually random and
symmetric on all scales below a given detector resolution, whatever the
current scale. As detector resolution increases, more and more detail is
revealed of the random walk which is statistically a self-similar fractal
(of dimension 2) on all scales. There is no inner characteristic scale in
this picture -- the mean free path of the particle is 0 in the continuum
limit. Similarly the mean free time is also zero, and the diffusive scaling
implies an infinite signal velocity in the continuum limit. All of these
features are evident in the resulting phenomenologies (the Diffusion
Equation and the Schr{\"{o}}dinger equation) for symmetric random walks and
entwined pairs alike.

In real physical diffusive systems, inner characteristic lengths, times and
velocities are all finite, and are determined by the density, composition
and temperature of the surrounding fluid. Three such measures are $l$, $\tau$
and $c$. $l$ is a mean free path and is roughly the average distance the
diffusing particle moves before being scattered by particles from the
surrounding medium. $\tau$ is roughly the expected time between scattering, $%
c$ is the ratio of these two and is roughly the speed of sound in the
system. For real physical systems then the scaling relation (\ref{limit}) is
a convenient mathematical fiction which allows us to replace a sequence of
difference equations with a limiting partial differential equation. The PDE
itself is then only a useful description on scales where the scaling
relation (\ref{limit}) is valid. For example, the diffusion equation is only
useful on space scales greater than $l$ and time scales greater than $\tau$.
Below these scales, a diffusing particle moves on a piecewise smooth path
rather than a Fractal trajectory, and the differential equation is no longer
a sensible description.

In quantum mechanics, there is a formal parallel to this in the transition
from Schr{\"{o}}dinger dynamics to the relativistic equations. The analog of
the characteristic speed of sound is of course the speed of light. The
analog of the mean free path is the Compton length. The analog of the mean
free time is the Compton time. All of this is well-known, and easily seen,
particularly in the path-integral formulation of quantum mechanics. However,
the reason for mentioning it here is that, for example, the relation between 
$\lambda _{C}$, the Compton length and $l$ the mean free path is purely
formal. $l$ is physically a measurable feature of a real classical system,
and mathematically an ensemble average. In contrast, in quantum mechanics $%
\lambda _{C}$ is a characteristic length at the level of the wavefunction
equations only. It is a physical parameter which is not directly measurable
as a distance between collisions, neither is it an ensemble average over any
known microscopic dynamic. The analogy between the two sets of parameters,
quantum and classical, is interesting but formal.

In this section we shall see that Entwined pairs generate the Dirac and Schr{%
\"o}dinger equation in such a way that the analogy between classical and
`quantum' characteristic lengths is no longer formal. $\lambda_C$ in the
context of entwined pairs is precicely a mean free path generated as an
ensemble average. The other characteristic constants are either prescribed
constants, ensemble averages, or, in the case of the `diffusion constant' $%
\hbar/(2m)$ an assumption on how space and time scale, as is the case for
the usual diffusion constant $D$.

Since equations (\ref{diff1}) and (\ref{diff2}) are very similar and are
coupled only by the initial conditions, we shall work with the first system
only. We use the scaling 
\begin{eqnarray}  \label{relscale}
\alpha&=&a\epsilon  \nonumber \\
\beta&=&(1-\alpha)\qquad\mathrm{and} \\
\frac{\delta}{\epsilon}&=&c  \nonumber
\end{eqnarray}
where $a$ and $c$ are both fixed parameters. $a$ is the inverse of the mean
free time of the system and $c$ is the fixed mean free speed. Substitution
of these into (\ref{diff1}) give. 
\begin{eqnarray}
\phi _{1}^{\prime }(z,t+\epsilon ) &=&(1-a\epsilon) \phi _{1}^{\prime
}(z+\delta,t )-a\epsilon\phi _{2}^{\prime }(z-\delta,t )  \label{diffD1} \\
\phi _{2}^{\prime }(z,t+\epsilon ) &=&a\epsilon \phi _{1}^{\prime
}(z+\delta,t )+(1-a\epsilon) \phi _{2}^{\prime }(z-\delta,t )  \nonumber
\end{eqnarray}

Notice that as $\epsilon\to0$, the likelihood of a direction change goes
down as $\epsilon$. Small steps are correlated in the same direction, giving
a finite mean free path.

Using the expansions of the shift operators (\ref{shiftx}) and (\ref{shiftt}%
) and truncating these to first order give 
\begin{eqnarray}
(1+\epsilon \frac{\partial }{\partial t})\phi _{1}^{\prime } &=&(1-a\epsilon
)(1+\delta \frac{\partial }{\partial z})\phi _{1}^{\prime }-a\epsilon
(1-\delta \frac{\partial }{\partial z})\phi _{2}^{\prime }+\mathrm{O}(\delta
^{2}) \\
(1+\delta \frac{\partial }{\partial t})\phi _{2}^{\prime } &=&a\epsilon
(1-\delta \frac{\partial }{\partial z})\phi _{1}^{\prime }+(1-a\epsilon
)(1+\delta \frac{\partial }{\partial z})\phi _{2}^{\prime }+\mathrm{O}%
(\delta ^{2})  \nonumber
\end{eqnarray}%
Matching first order terms and using $\delta =\epsilon c$ gives: 
\begin{eqnarray}
\frac{\partial \phi _{1}^{\prime }}{\partial t} &=&c\frac{\partial \phi
_{1}^{\prime }}{\partial z}-a\phi _{2}^{\prime }  \label{diracq} \\
\frac{\partial \phi _{2}^{\prime }}{\partial t} &=&-c\frac{\partial \phi
_{2}^{\prime }}{\partial z}+a\phi _{1}^{\prime }  \nonumber
\end{eqnarray}%
Examining (\ref{diracq}) we can see that the $\phi $ are real, but
oscillatory. The oscillatory character arises through the two different
signs in the scattering terms on the right of the equation. The alternating
signs in these terms are a result of the entwining of the paths and the
resulting `colouring' of the envelopes(see fig. 1). This is the origin of
`phase' in this system. Note that if we start with initial conditions such
that the $\phi ^{\prime }$ are constant in space, the spatial derivatives
are zero and the system reduces to 
\begin{eqnarray}
\frac{\partial \phi _{1}^{\prime }}{\partial t} &=&-a\phi _{2}^{\prime }
\label{diracq0} \\
\frac{\partial \phi _{2}^{\prime }}{\partial t} &=&a\phi _{1}^{\prime } 
\nonumber
\end{eqnarray}%
A suggestive solution of (\ref{diracq0}) is: 
\begin{eqnarray}
\phi _{1}^{\prime } &=&A\cos (at) \\
\phi _{2}^{\prime } &=&A\sin (at)  \nonumber  \label{homosol}
\end{eqnarray}%
The trigonometric functions signal the implicit presence of phase in the
system. Notice that were we only counting paths without the return path
present, the sign of the scattering terms in (\ref{diracq}) would both be
positive and the solutions (\ref{diracq0}) would be hyperbolic, not
trigonometric.

With the appropriate numerical constant for $a$,(ie. $a=mc^2/\hbar$)
equation (\ref{diracq}) is a form of the Dirac equation in one dimension
which admits real solutions. Since in this context the constant $c$ is the
speed of the particle on the lattice, which has been fixed in the continuum
limit, we should be able to recover the `non-relativistic' limit by letting $%
c\to \infty$. To do this we shall rewrite equations (\ref{diracq}) in a more
convenient form. First we change variables to: 
\begin{eqnarray}
\psi _{1}&=&i\phi _{1}^{\prime } e^{-iat}  \nonumber \\
\psi _{2}&=&\phi _{2}^{\prime } e^{-iat}  \label{subst1}
\end{eqnarray}
The $i$'s in (\ref{subst1}) are not FAC's. They simply give a convenient
linear combination of the two densities $\phi^{\prime}$. Equation (\ref%
{diracq}) then becomes

\begin{eqnarray}
\frac{\partial \psi _{+}}{\partial t} &=&c\frac{\partial \psi _{-}}{\partial
z}  \label{diracL1} \\
\frac{\partial \psi _{-}}{\partial t} &=&c\frac{\partial \psi _{+}}{\partial
z}+2ia\psi _{-}  \label{diracL2}
\end{eqnarray}%
Now eliminate $\psi _{+}$ by differentiating (\ref{diracL1}) with respect to 
$z$ and (\ref{diracL2}) with respect to $t$ and combining to get 
\begin{equation}
\frac{1}{c^{2}}\frac{\partial ^{2}\psi _{-}}{\partial t^{2}}=\left( \frac{2ia%
}{c^{2}}\right) \frac{\partial \psi _{-}}{\partial t}+\frac{\partial
^{2}\psi _{-}}{\partial z^{2}}.  \label{schrod2}
\end{equation}%
In this form, contact with the Schr{\"{o}}dinger equation is easily made.
The choice of $a$ which identifies (\ref{diracq}) with the usual Dirac
equation is $a=mc^{2}/\hbar $. So $a$ depends on $c$. If we substitute this
into (\ref{schrod2}) we get 
\begin{equation}
i\frac{\partial \psi _{-}}{\partial t}=-\left( \frac{\hbar }{2m}\right)
\left( \frac{\partial ^{2}\psi _{-}}{\partial z^{2}}-\frac{1}{c^{2}}\frac{%
\partial ^{2}\psi _{-}}{\partial t^{2}}\right) .  \label{newschrod}
\end{equation}%
where now the only term that depends on $c$ is the last one. Here, as
expected, $\lim_{c\rightarrow \infty }$ gives the free particle Schr{\"{o}}%
dinger equation. Furthermore, if we take $m$ to the left hand side of (\ref%
{newschrod}) we see that in the limit as $m\rightarrow 0$, $\psi _{-}$ obeys
the wave equation. This makes sense from the original model in that when $m=0
$, the particles never scatter, so they stay on the same light-cone and
hence they obey the wave equation. By writing the solutions of (\ref%
{newschrod}) in the form $\psi _{-}=\exp (imc^{2}t)\chi $ it is seen that (%
\ref{newschrod}) is equivalent to the Klein-Gordon equation for a particle
of mass $m$. 

Note that the above route to the Schr{\"o}dinger equation did not include a
potential; we arrived only at the free particle equation. To include a
potential in this case would require us to put field interactions into the
Dirac equation (\ref{diracq}) which are not real. This may be done, but
would be difficult to interpret in terms of the original stochastic model.
Instead,in a subsequent paper we shall show how to include a field, itself
generated by a classical stochastic process, which will allow us to avoid
this second FAC.

The two approaches above represent two different methods of taking a
continuum limit. In the first approach we took a single continuum limit in
which, as the lattice spacing went to zero, the speed of the particle on the
lattice increased without bound. This can be seen from the diffusive scaling
in (\ref{limit}) which can also be written.

\begin{equation}
(\delta )/(\epsilon )\,\rightarrow 2D/\delta \qquad \mathrm{{as}\qquad }%
\epsilon ,\delta \rightarrow 0  \label{speed}
\end{equation}%
where $\delta /\epsilon $ is just the hopping speed of the particle on the
lattice. This is the appropriate scaling for symmetric random walks and
reflects their intrinsic Fractal dimension. In the context of Schr{\"{o}}%
dinger 's equation it is this scaling that supports the uncertainty
principle. Note that if we call the lattice speed as a function of the
lattice scale $v(\delta )$, (\ref{speed}) may be suggestively written: 
\begin{equation}
v(\delta )\times \delta \approx 2D
\end{equation}%
The apparent speed of a particle increases as the scale of measurement goes
down in such a way that the product of the length scale and speed is
asymptotically constant. This in turn would put a lower bound on the product
of the uncertainties in the spatial resolution and speed which would depend
only on D.

In the second route to Schr{\"o}dinger 's equation we acknowledged that
there should be a finite signal velocity $c$. This meant that the above
scaling could only hold down to some characteristic length which in our
model was $1/a$. Below that scale, particle speeds had to be constant and as
a result the continuum limit was taken with the scaling (\ref{relscale}).
This scaling yields trajectories in the continuum which look like the paths
in Fig. 1, except there is no lattice, and the lengths of the line segments
between corners is governed by a Poisson process, with an expected length of 
$1/a$. When the continuum limit is taken in this fashion, the ultimate
particle speed $c$ remains a parameter in the resulting PDE. By expressing
the PDE in a convenient form we  then let $c\to \infty$ to recover `the
non-relativistic' form of the equation. The conceptual link with the first
continuum lmit is that our broken line paths look like ordinary symmetric
random walks if the scale of measurement in space and time has a natural
`speed' much less than the ultimate speed $c$. By sending $c \to \infty$ you
ensure that this is always the case, and the resulting PDE is the same for
both cases.

\section{Discussion}

In the above, we showed that in 1+1 dimensions, the Schr{\"{o}}dinger
equation arises as the continuum limit of a classical stochastic model by
taking the limit of a discrete system in two separate ways. 
%

In qualitative terms, the above calculations replaced the Brownian motion
underlying the diffusion equation by the Brownian motion of entwined paths. 
This replacement has then changed the macroscopic phenomenology from
the Diffusion equation to the Schr{\"o}dinger equation. The `anti-particle'
current of the entwined paths provides the interference effects and
reversibility characteristic of Schr{\"o}dinger's equation.   

The idea that Schr{\"{o}}dinger 's equation somehow involves time-reversed
`fluids' has been the source of many alternative approaches to quantum
mechanics\cite{Nelson66,Nelson85,Nagasawa96,Nottale92,Naschie95c}. Our work
can be understood as a modification of Brownian/Poisson motion that provides
a microscopic basis (and resulting statistical mechanics) for a
time-symmetric fluid. It thus provides a new context for the Schr{\"{o}}%
dinger equation as a legitimate phenomenological equation for entwined
paths. This is in marked contrast to quantum mechanics where the Schr{\"{o}}%
dinger equation is the fundamental equation of the theory.

In terms of conventional single-particle quantum mechanics in one dimension,
the above model imitates it exactly, up to the measurement postulates, in a
realistic context. In terms of the Penrose\cite{Penrose89} partition of
quantum mechanics into `U' (unitary) processes and `R' (reduction)
processes, Entwined pairs provide a microscopic model for the U-process.

 It
is tempting to classify the model as a `hidden variables model'. However
this would be misleading. Entwined paths are to the Schr{\"{o}}dinger
equation what ordinary random walks are to diffusion. Few people would call
Brownian motion a hidden variable theory for Diffusion. However if we graft
the measurement postulates onto the `wavefunctions' generated by entwined
paths, then we do indeed have a (non-local) hidden variables theory.
However, it makes little sense to graft the measurement postulates onto a
realistic theory. Just as we do not need measurement postulates for
diffusive systems, we do not need postulates for entwined paths. Instead,
the task ahead is to see if any reasonable measurement schemes verify or
contradict the postulates of quantum mechanics. By such tests we stand to
gain insight into how wave-particle duality might, or might not, be
produced in Nature.


\begin{acknowledgments}
GNO is grateful for many helpful discussions with J.A. Gualtieri. This work
was partly supported by the Natural Sciences and Engineering Research
Council of Canada.
\end{acknowledgments}  


\end{document}